\documentclass[]{mn2e}

\usepackage{amssymb}
\usepackage{graphicx}
\usepackage{amsmath}
\usepackage{graphics}

\usepackage{color}

\title[Diffuse Galactic IR emission]{The contribution of CHONS particles to the diffuse high Galactic latitude IR emission}
\author[R. Papoular]{R. Papoular$^{1}$\thanks{E-mail:
papoular@wanadoo.fr}\\
$^{1}$Service d'Astrophysique and Service de Chimie Moleculaire,\\
CEA Saclay, 91191 Gif-s-Yvette, France}

\begin{document}

   \maketitle
\label{firstpage}

\begin{abstract}
This work purports to model the far infrared gray-body emission in the spectra of high-Galactic-latitude clouds. Several carbonaceous laboratory materials are tested for their fitness as carriers of this modified-black-body emission which, according to data delivered by the \emph{Planck} satellite, and others before, is best fit with temperature 17.9 K and spectral index $\beta$=1.78. Some of these materials were discarded for insufficient emissivity, others for inadequate $\beta$. By contrast, CHONS clusters ($\beta=1.4$, $T=19$ K) combine nicely with magnesium silicate ($\beta=2$, $T=18.7$ K) to give a spectrum which falls well within the observational error bars (total emission cross-section at 250${\mu}$m: $8.6\,10^{-26}$ cm$^{2}$ per H atom). Only 15 \% of all Galactic carbon atoms are needed for this purpose. The CHONS particles that were considered and described have a disordered (amorphous) structure but include a sizable fraction of aromatic rings, although they are much less graphitized than a-C:H/HAC. They can be seen as one embodiment of ``astronomical graphite" deduced earlier on from the then available astronomical observations.

Grain heating by H atom capture is proposed as a contributor to the observed residual emissions that do not follow the dust/HI correlation.

\end{abstract}




\section{Introduction}
The space mission \emph{Planck} (Tauber et al. 2010, Planck collaboration 2011) has observed the sky in a wide frequency range, from the cosmic microwave band to the FIR (Far InfraRed) with high sensitivity and angular resolution. In particular, the bolometers of its High Frequency Instrument, cooled to 0.1 K, cover the 100, 143, 217, 353, 535 and 857 GHz bands. They delivered exquisitely detailed and accurate measures of electro-magnetic emission from the Diffuse Galactic Interstellar Medium (DGISM) at high latitudes, which allowed the probing of the full Spectral Energy Distribution (SED) of the thermal emission of large dust grains that make up a sizable fraction of the InterStellar (IS) dust mass. These results complement and support previous space missions which contributed considerably to the same field, such as the Infra Red Astronomical Satellite (IRAS) and the Cosmic Background Explorer (COBE/FIRAS).

The present work draws heavily, in particular, on the analyses of these measurements made by Boulanger et al. \cite{bou96}, Abergel et al. \cite{abe11} and Compiegne at al. \cite{com11}. These authors established that the average thermal spectrum of thin, quiescent clouds in the local DGISM could be fitted by a gray-body spectrum at $T_{th}=17.9\pm \,1\,$K, with an emission cross-section having a power law form with exponent (spectral index) $\beta=1.78\pm0.18$ and \bf a value of the emission/absorption cross-section \rm  $\sigma_{0}=1\pm0.3\,10^{-25}$ cm$^{2}$ per H atom at 250 $\mu$m. For the record, we note that Ade et al. \cite{ade11}, studying the solar neighborhood, find 
 $\beta=1.8\pm0.18$ and $\sigma_{0}=0.52\,10^{-25}$ cm$^{2}$ per H atom at 250 $\mu$m.
 
 Given the presence, in the ISM, of materials with different 
 physical structures and chemical composition (carbonaceous and silicious), embedded in gas of various degrees of dissociation/ionization, such a pure, not composite, fit is remarkable. In fact, Fig. 17 of Abergel et al. \cite{abe11} hints at the extent of \bf deviations \rm from the best fit, for clouds of different velocities, at different sites. Earlier on, Paradis et al. \cite{par} also found evidence of considerable variations in spectral index, from 0.9 to 2.8, depending on site and wavelength.

It is shown in Sec. 2 below that, in fact, the observational error bars allow for small  anti-correlated excursions of $T_{th}$ and $\beta$. Some variety of materials and ambient conditions can thus be reconciled with the data.
 
 The question is then addressed, in Sec. 3, of which materials comply with the observational constraints. This, of course, is an age-old problem, which has already been given many answers, in particular by Mathis et al. \cite{mat77}, Mezger et al. \cite{mez82},  Mathis et al.\cite{mat83}, who selected silicates and graphite. Pollack et al. \cite{pol} invoked refractory materials (inspired by the kerogens and tholins studied by Khare et al. 1990), for which they give power indexes; the latter were also used by Finkbeiner et al.\cite{fin}. For the far IR, Draine and Li \cite{dra07} used amorphous silicates and graphite. More recently, so-called "amorphous carbon" with the properties of BE (benzene burned in air) extracts was used by Compiegne at al. \cite{com11} to model the observed spectrum; Abergel et al. \cite{abe13} appear to have made use of Pollack's extinction data.
 
 The initial, celebrated MRN model of the former authors, however, was specifically tailored for the spectral range from 1 to 0.1 $\mu$m. It had, therefore, to be extrapolated into the FIR, a step which Mathis himself considered perilous (Mathis and Wallenhorst 1981). Indeed, the astronomical data available at that time were much less constraining than the \emph{Planck} data. On the other hand, laboratory experiments and measurements on various, well-characterized, materials have since been accumulating, especially on graphite, which incites a revisiting of this issue. 

Moreover, beyond 10 $\mu$m and at low temperatures, it does not seem that particle size is as relevant as in the vis/UV, where it performed an essential role in the MRN model (see Draine and Lee 1984, Fig. 4).

Furthermore, for grains to be in thermal equilibrium, their cross-section must be large enough as to capture enough radiative energy during their radiative cooling time. For the DGISM, this implies sizes over about 1 $\mu$m (see Draine and Anderson 1985), which is outside the MRN distribution.

In the present work, the emphasis is on more or less large molecules rather than on more or less small solid-state grains. Modern modeling softwares deliver directly the integrated absorption efficiency of a molecule, small or large, and dispenses with the application of Mie's theory with its complicated dependence on size and shape. This approach also allows one to explore the quality, as carrier candidates, of materials other than graphite and silicates: we shall consider here, in Sec. 4, various embodiments of the ubiquitous CHONS particles, so called because they are made of some of the most abundant elements in the Galaxy (H, C, O, N and S). These molecules are constitutive of the organic refractory matter called kerogen, that is found at shallow depths in earth, and were first invoked by planetologists in the study of meteorites (see Khare et al. 1990). More recently, Kwok \cite{kwo} suggested that a similar mixture of disorganized aromatic/aliphatic structures were synthesized in the circumstellar environments of stars in their late stages of evolution; he designated them under the name MAONS (Mixed Aromatic/aliphatic Organic Nanoparticles). Such mixtures will be shown below to satisfy the requirements for them to contribute to the DGISM emission, with no tight constraints on size or shape.

Section 5 then shows how CHONS and silicate particles can be combined to emit a gray-body spectrum that falls within the error bars of \emph{Planck}'s measurements with an emission cross-section of $0.86\,10^{-25}$ cm$^{2}$ per H atom at $\lambda$=250$\,\mu$m.

The very definition of an emission cross-section as per H atom implies that the excitation agent is independent of the atomic density of the medium, which is the case of the ambient UV radiation in thin media. Abergel et al. \cite{abe11}, however, found important deviations from this condition (``residuals") and proposed Cosmic Background Radiation irregular interference to be their main cause. Section 6, below, discusses the possibility that grain heating by H atom capture might contribute to those discrepancies.

The paper closes with a Discussion of the present findings.

\section{A range of observational gray-body parameters}
\begin{figure}
\resizebox{\hsize}{!}{\includegraphics{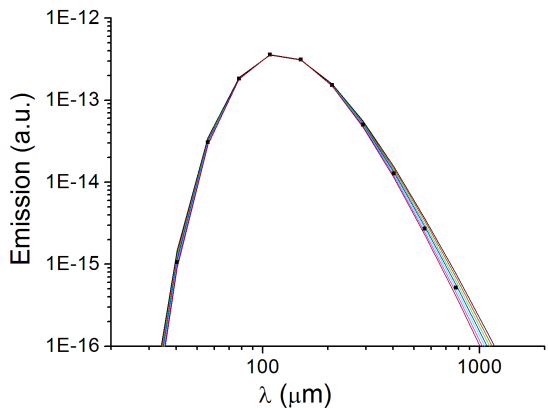}}
\caption[]{Best fit spectra to the FIRAS/\emph{Planck} data, for different spectral indexes, $\beta$. The spectral shape is here represented by the quantity $F=cBB(T,\lambda)\lambda^{-\beta}$, drawn as lines for arbitrary values of $\beta$: 1 (black), 1.5 (red), 1.6 (green), 1.7 (blue), 1.8 (cyan), 1.9 (magenta) and 2 (wine). The corresponding temperatures, $T$, and constant multiplying factors, $c$, are obtained by fitting $F$ to the best-fit SED of Abergel et al. \cite{abe11}, which is represented by black squares. These temperatures are plotted in Fig.\ref{Fig:correlation}. \bf Figure \ref{Fig:residuals} helps evaluating the small differences between the curves drawn here. \rm}
\label{Fig:graybody}
\end{figure}

The spectral shape of a gray body is determined by the product $BB(T,\lambda)\lambda^{-\beta}$,
where $BB$ is the black-body law. This quantity is plotted in Fig. \ref{Fig:graybody} (black squares) for a gray-body which best fits the observations of the DGISM at high latitudes (Abergel et al. 2011, Fig. 17). This gray body has 

$T=17.9 \,\mathrm{K} ,  \,\beta=1.78$.

\begin{figure}
\resizebox{\hsize}{!}{\includegraphics{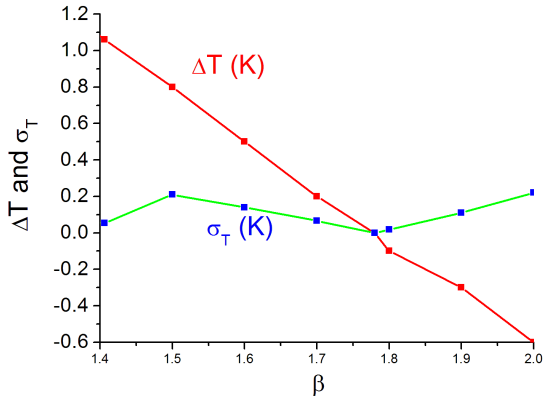}}
\caption[]{In red, the temperatures which best fit the quantity $BB(T,\lambda)\lambda^{-\beta}$ to the best-fit SED of Abergel et al. \cite{abe11}. In blue squares and green line, the corresponding standard errors of the fits.}
\label{Fig:correlation}
\end{figure}

\begin{figure}
\resizebox{\hsize}{!}{\includegraphics{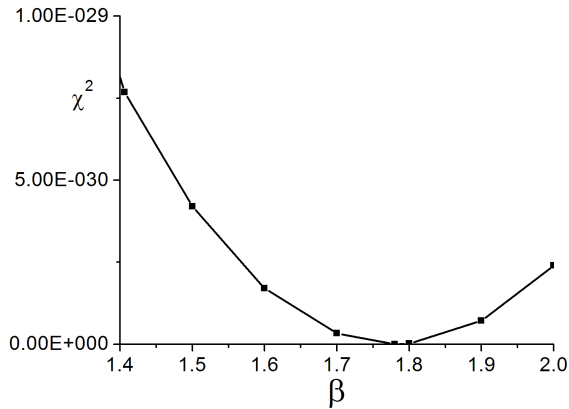}}
\caption[]{The $\chi^{2}$ statistical estimator corresponding to the cases illustrated in
Fig.\ref{Fig:graybody}.}
\label{Fig:chisquare}
\end{figure}

\begin{figure}
\resizebox{\hsize}{!}{\includegraphics{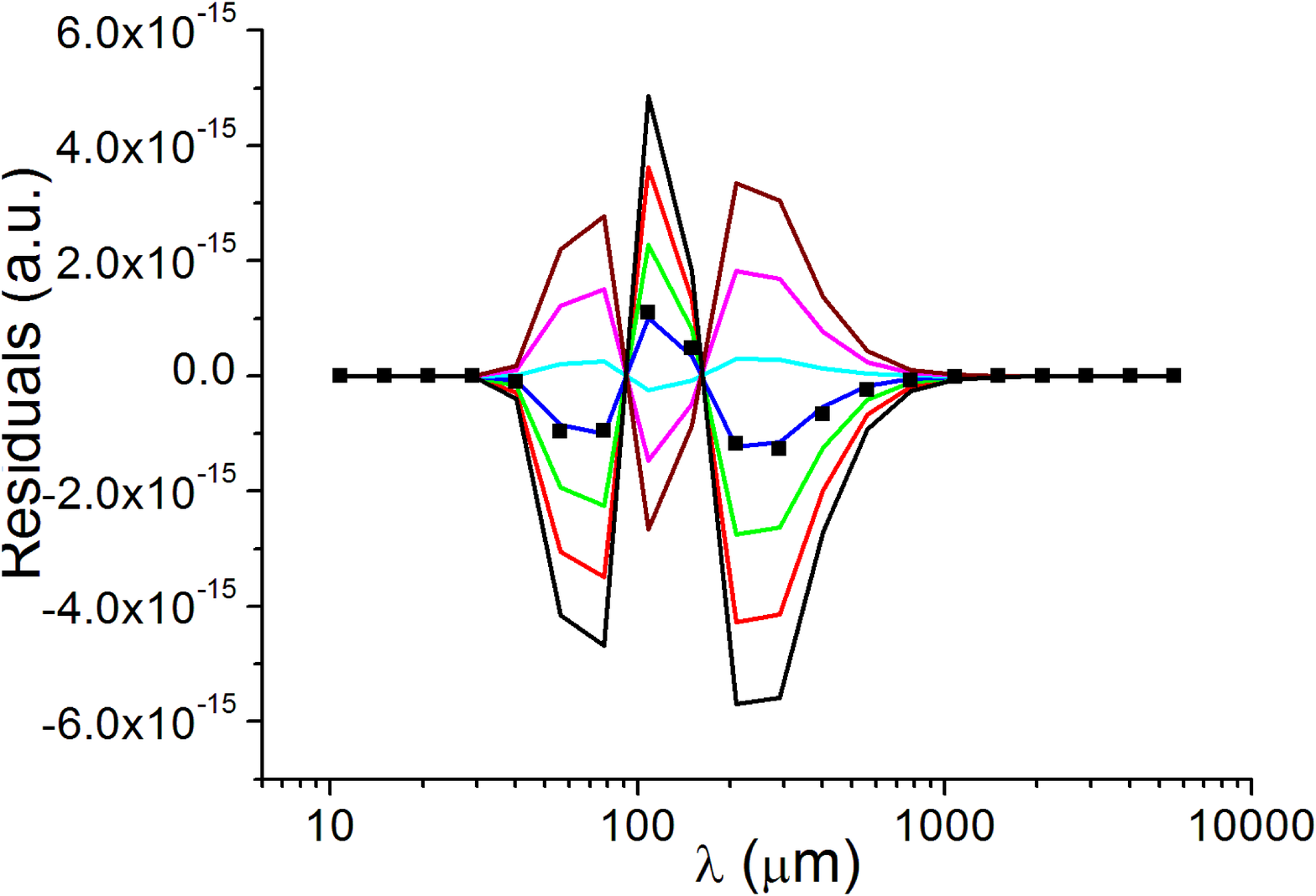}}
\caption[]{The residuals $r=F_{0}-F$, where $F=cBB(T,\lambda)\lambda^{-\beta}$, for $\beta=$1.4 (black), 1.5 (red), 1.6 (green), 1.7 (blue), 1.8 (cyan), 1.9 (magenta) and 2 (wine); average: black squares. \bf In a way, this may serve as a zoom on the small differences between the curves in Fig. \ref{Fig:graybody}. \rm}
\label{Fig:residuals}
\end{figure}

 Superimposed upon it are fits to it using the function $cBB(T,\lambda)\lambda^{-\beta}$, where
 $c$ and $T$ are the fitting parameters and $\beta$ is successively constrained to discrete values in the range known to encompass several common carbonaceous materials (1 to 2). The corresponding values of $\Delta T=T_{fit}-17.9$ and the attendant standard errors, $\sigma_{T}$, are displayed in Fig.\ref{Fig:correlation}. The $\chi^{2}$ and the residuals of the fits are shown in Fig.\ref{Fig:chisquare} and Fig.\ref{Fig:residuals}. Note the large ranges of allowable anti-correlated values of $\beta$ and $T$. 
 
As expected, Fig.\ref{Fig:residuals} shows that, for $\beta<1.78$ (respectively $\beta>1.78)$, the Spectral Energy Distribution (SED)is slightly widened (respectively narrowed) and shifted downward (upward) with respect to the best fit. As a consequence, the average residual for an equal-weight combination of $\beta=$1.4 and 2, for instance, is mostly reduced with respect to each separately, as illustrated by the black squares.

The error bars in Fig. 17 of Abergel et al.\cite{abe11} are due to the dispersion of site properties. Obviously, they give enough leeway to accommodate a large gamut of materials including silicates and graphite, which have high $\beta$, as well as some amorphous carbon, which have low $\beta$. That is, of course, assuming they also nearly satisfy the condition on absolute value of the emission/absorption cross-section (Abergel et al. 2011)
\begin{equation}
\sigma(250\,\mu\mathrm{m})=1.0\pm0.3\,10^{-25} \mathrm{cm}^{2}\mathrm{per\,H\,atom}.
\end{equation}

 For grains small relative to the wavelength, this cross-section (c-s) is linked to their optical c-s, $C$, by

\begin{equation}
\sigma=f\Lambda\frac{C}{N_{E}},
\end{equation}

where $f$ is the fraction of active element, $E$, tied up in grains, $\Lambda$ is the cosmic abundance of this element, and $N_{E}$ the number of atoms of this element in each grain. Assuming the grains to be approximately spherical, with radius $a$, this relation becomes
 
\begin{equation}
\sigma=f\Lambda\frac{Q/a}{n_{E}},
\end{equation}

where $Q$, is the extinction efficiency, and $n_{E}$ the number density of atoms $E$ in the grain. The corresponding expression for molecules is derived below.

Now, in many cases of interest here, the model material samples to be measured come in pressed pellets of small grains and the extinction is given in terms of $\alpha$ cm$^{-1}$, or $\kappa$  cm$^{2}$g$^{-1}$, which is $\alpha$ divided by the density of the material in g.cm$^{-3}$. As the optical constants are usually not given, it is not possible to compute $Q/a$. However, because of the high pressures applied, it is usually nearly the case that the number density of $E$ atoms in the pellet is equal to $n_{E}$; then 

\begin{equation}
Q/a\sim\alpha.
\end{equation}

In order to accommodate various cases in the same graphs for comparison, we use these relations to express all given extinctions in terms of $\alpha$. 

Considering carbon-rich models first, take $\Lambda=1/3000$ and $n=10^{23}$, corresponding to a typical solid state density, $\approx2$ g.cm$^{-3}$; also assume, for a moment, that the C atoms are equally distributed among CO$_{2}$ molecules and dust; then $\sigma_{C}=\alpha/6\,10^{26}$. If, initially and tentatively, carbonaceous grains are only required to provide half the observed emission power, the other half being provided by the silicate grains, then the best fit value for $\alpha$ is approximately 30 cm$^{-1}$ at $\lambda=250\,\mu$m. 

\begin{figure}
\resizebox{\hsize}{!}{\includegraphics{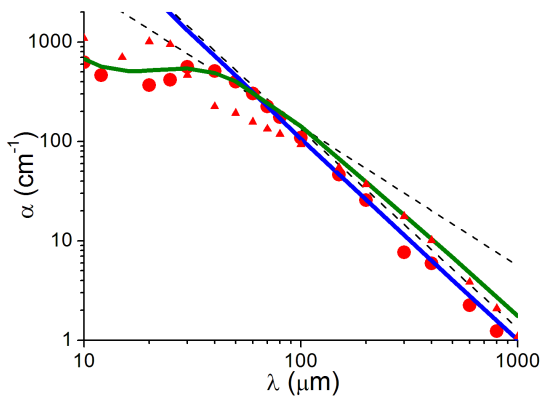}}
\caption[]{\it Dotted lines: \rm lines with logarithmic slopes 1.4 and 2, respectively passing through $\sigma=1.3\,10^{-25}$ and $0.7\,10^{-25}$ cm$^{2}$ per H atom at 250 $\mu$m ($\alpha=30(1\pm0.3)$  cm$^{-1}$, respectively), so as to bracket the \emph{Planck} best fit with its error bars. Best model fits to observations over the years: \it Red dots and triangles:\,\rm  Mathis et al. \cite{mat83} for ``astrophysical" graphite and silicate, respectively. \it Thin blue line:\,\rm   Draine \cite{dra85} for ``astrophysical" graphite. \it Thick olive line\rm:  Draine \cite{dra85} for ``astrophysical" silicate. }
\label{Fig:alpha1}
\end{figure}

Figure \ref{Fig:residuals} indicates that, if the residuals are to be kept below $\approx\Longleftarrow 1\%$ \bf of the maximum in Fig. \ref{Fig:graybody}, \rm then $\beta$ should lie between $\approx$1.4 and 2. Figure \ref{Fig:alpha1} is meant to encapsulate these  constraints on the extinction $\alpha$ of candidate model carriers:
the dotted lines are drawn with these logarithmic slopes, but they are respectively shifted upwards and downwards by 30$\%$ to take into account the uncertainty in $\sigma_{0}$, and, therefore, on $\alpha$. Superimposed upon these lines are curves representing data from Mathis et al. \cite{mat83} and Draine \cite{dra85}, which were fitted to older observations. Note that the silicate curve from Mathis et al. \cite{mat83} was shifted downward by a factor $\sim$10 to account for the nearly 10-fold smaller cosmic abundance of silicon relative to that of carbon.

It is remarkable how close the data of these authors come to those defined by the FIRAS/\emph {Planck}  measurements. However it was not claimed that the corresponding so-called ``astrophysical" properties were indeed those of a terrestrial material, especially for ``graphite" (which was tailored mainly to fit the 2175 $\AA{\ }$ extinction feature). Indeed, the following shows that, at least for graphite, this is not the case.

\section{A range of candidate models}

A number of natural or laboratory materials have been considered as candidate carriers for the IR continuum. Some are reviewed here in the light of the FIRAS/\emph{Planck} constraints.

\begin{figure}
\resizebox{\hsize}{!}{\includegraphics{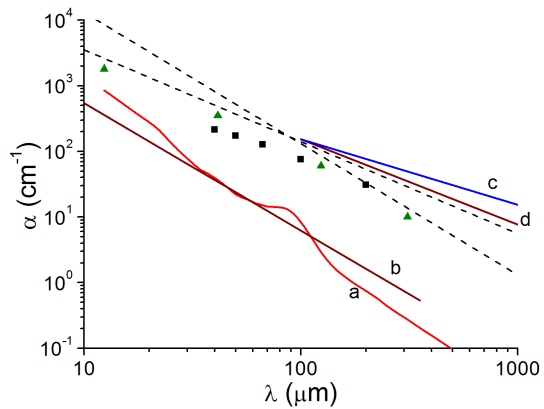}}
\caption[]{\it Dashed lines: \rm lines with logarithmic slopes 1.4 and 2, as in 
Fig. \ref{Fig:alpha1}. (a) Extinction of graphite from Philipp's dielectric functions \cite{phi77}. (b) Graphite extinction from electronic band theory (Yuan et al. \cite{yuan11}). (c)  Laboratory amorphous carbons TU, BE, XY (shifted downward by $\times$0.19); adapted from Koike et al. \cite{koi94}. (d) One of the lesser evolved (graphitized) coals, adapted from Mennella et al. \cite{men95} (shifted downward by $\times$0.51). \emph{Green triangles}: glassy carbon, adapted from Rouleau and Martin \cite{rou}. \emph{Black squares}: laboratory amorphous magnesium silicate, adapted from Day \cite{day76}, $\times$0.103. See text.}                  

\label{Fig:alpha2}
\end{figure}

The most popular data on graphite in the FIR are those provided by the measurements and analysis of Philipp \cite{phi77}. The extinction efficiency and, hence, $\alpha$ can be obtained from the dielectric constants he deduced from his measurements. This is plotted as curve (a) in Fig.\ref{Fig:alpha2}, in which the best-fit limits are drawn as dashed lines, as in Fig. \ref{Fig:alpha1}. Although it has an acceptable slope, it is too weak by about a factor 100. As an aside, note that the previous reflectance measurements of Taft and Philipp \cite{taf65} did not imply the existence of the hump near 80 $\mu$m (see Philipp 1977, Fig. 1); neither do more recent reflectance measurements (Kuzmenko et al. 2008, Fig. 1 and 2).
 
Curve (b) in the same figure plots the result of a  recent tight-binding calculation of graphene and graphite by Yuan et al. \cite{yuan11} (see Papoular et al. 2013 for details). Except for the absence of the hump around 80 $\mu$m, this is quite close to the Philipp curve 
(both are for $\vec {E}\perp \vec{c}$). Thus, terrestrial graphite can hardly compete as a carrier model for the FIR.

Pursuing, curve (c) represents the average behavior of amorphous carbons prepared in the laboratory by Koike et al. \cite{koi94}, down shifted by a factor 0.19 to pass through the crossing point of the dashed lines. While this indicates a strong FIR efficiency, the average slope is too low, at $\sim1$. Similarly, curve (d) plots the extinction of a weakly graphitized coal from the Mericourt (France) mine, as measured by Mennella et al. \cite{men95} but shifted downwards by a factor 0.51. With a logarithmic slope of 1.3, and a strong extinction, this comes closer to the \emph{Planck} requirements. The other materials measured by Mennella et al. \cite{men95} have $\beta$=0.9, 0.9 and 0.8 for BS (powder of $\beta$-silicon carbide), AC (soot from an arc discharge in argon) and BE (soot from benzene burned in air) respectively, and do not comply with the constraints. (As the data in Fig. 11 of Mennella et al \cite{men95} are in cm$^{2}$.g$^{-1}$, the $\alpha$'s, here, were deduced assuming the material density to be 2 g.cm$^{-3}$).  The amorphous carbon model used by Compiegne et al. \cite{com11} is also basically derived from the measurement of Mennella et al \cite{men95} on the same BE sample, but they use $\beta=1.55$ for the absorption spectral index (see their Fig. A1). This may be linked to the intermediate step they took, consisting in first deriving the corresponding optical refraction indexes (which were required for their Mie calculations).

Rouleau and Martin \cite{rou} had previously considered the effects of shape, clustering and porosity on AC and BE. Their computations are in rough agreement with the measurements of Mennella et al. \cite{men95}. Besides, they also considered glassy carbon, both in CDE (Continuous Distribution of Ellipsoids) and spheres. The corresponding extinction slopes were found to be about 0.9 and 1.4 respectively; the latter is the strongest slope in their study, and is represented by green triangles in Fig.\ref{Fig:alpha2}. This is clearly the most fitting material among those found in the literature.  

Glassy (or vitreous) carbon has a simple composition (only carbon) and a complex structure. The bonds are believed to be nearly 100$\%$ $sp^{2}$ but the longer-range order is not graphitic; rather it may be represented by entangled fibrils (see Robertson 1986, Fig. 3a). Obviously, this type of structure is not likely to be produced in large quantities in low density space. Moreover, the absolute values of extinction of glassy carbon imply that all the carbon atoms available in condensed dust should be built into such a structure, which is still less likely.

In general, strong electronic absorption can only be expected from a metal, and not too cold at that. Graphite is only a semi-metal and its conductivity in the spectral range of interest only becomes significant above 0 K, and if it is doped with impurities so as to let some electrons free. That is why hydrogenated carbon is distinctly more efficient. Glassy carbon, on the other hand, has a stronger extinction because of its fibril or lamellar structure. Other metals, such as iron, are too sparse in space (see Huffman 1977, Fig. 48).

On the other hand, the IR is known to be the domain of atomic vibrations. There is no better illustration, in the present context, than the spectrum of silicates such as Mg$_{2}$SiO$_{4}$ 
and Fe$_{2}$SiO$_{4}$, which are believed to be born around RGB stars. In 
Fig.\ref{Fig:alpha2}, the \bf black \rm squares represent measurements on a laboratory-produced amorphous magnesium silicate adapted from Day \cite{day76}, Fig. 3.3. However, in order for silicates to be compared to the other materials and to the \emph{Planck} constraints \bf on the FIR slope of $\alpha$ \rm, it was down-shifted by a factor \bf 0.103 \rm to account for the weaker silicon cosmic abundance. Obviously, if it is also assumed that silicates are only required to provide half the total astronomical extinction, 
the vibrational absorption of laboratory silicate is close to complying with \emph{Planck}'s constraints. It is reasonable, then, to explore the \emph{vibrational} spectra of amorphous carbonaceous materials as well, which is done below.

 \section{The FIR vibrational spectrum of a small grain}
 
The present study of vibrational spectra starts here with molecules as opposed to solid samples. For a solid, $\alpha$ is a continuous function of the solid's dielectric functions or optical indexes, which are also continuous. Molecular spectra, on the other hand, are discrete line spectra, and $\alpha$ \it for a single line \rm is deduced from the corresponding integrated band absorption, $A$, by

\begin{equation}
\alpha(\lambda)=10^{2}A(\lambda)\frac{C}{\Delta\nu}\,\,,
\end{equation}

where $\alpha(\lambda)$ is in cm$^{-1}$, $C$ is the molecular density (mol.l$^{-1}$) along the line of sight, and $\Delta\nu$ is the band width (cm$^{-1}$), which depends on the environment 
(see Atkins 2004). $A$ is obtained from a calculation of the normal modes of vibration of the molecule, including mode frequency, electronic energy distribution, and velocities of motions of all atoms in each mode. This gives the electric polarizability and, hence, $A$ then $\alpha$.
It is important to note that, by construction, these values take into account the size and conformation of the structure, so that \it the notion of surface modes invoked for very small solid grains is unnecessary here.\rm

\begin{figure}
\resizebox{\hsize}{!}{\includegraphics{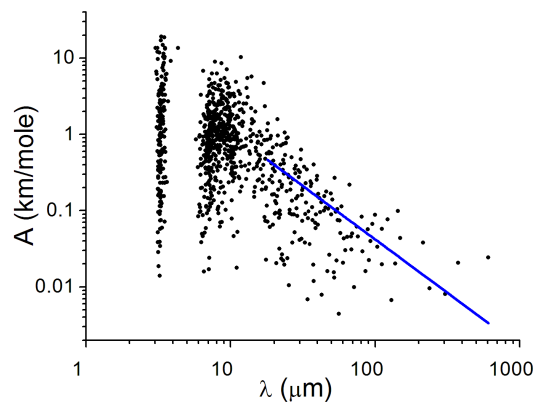}}
\caption[]{The integrated absorption line spectrum of a model structure for a-C:H, consisting of 110 C atoms and 182 H atoms interconnected by essentially \it sp\rm $^{1}$ and \it sp\rm $^{3}$ bonds. Each vibrational mode is represented by a point. Note the strong dispersion of modes in wavelength and intensity. The line represents an average of intensities over the phonon range; its logarithmic slope is 1.4. The mode density scales like $\lambda^{-2.6}$.}
\label{Fig:hacspec}
\end{figure}

An example of spectrum is given in Fig.\ref{Fig:hacspec} for the case of an a-C:H structure. This term is used here to designate a disordered structure made up of only carbon and hydrogen atoms, interconnected by essentially \it sp\rm $^{1}$ and \it sp\rm $^{3}$ bonds as opposed to other amorphous carbons, which are often more aromatic. It was obtained with a commercial chemical modeling software as described in Papoular \cite{pap13b}. The software performs automatically the calculations described above and, in addition, provides visual description on screen of the atomic motions in each individual mode.

In Fig.\ref{Fig:hacspec}, each vibrational mode is represented by a point with coordinates $\lambda$ and $A$, so as to emphasize the strong dispersion of modes in wavelength and intensity. Their number is 3N-6, where N is the total number of atoms in the structure. Inspection of the atomic displacement distribution on the screen, for a given mode, immediately reveals the fingerprint modes, whose vibrations are spatially limited to chemical functional groups; they do not extend beyond $\sim20\,\mu$m in wavelength. A large fraction of the modes extend well beyond and are characterized by unlocalized atomic vibrations all over the structure.
They never occur below 5$\,\mu$ and their upper wavelength limits increases with molecular size.
Their spectral density decreases steeply as $\lambda$ increases beyond about 18 $\mu$m, as the figure shows. This behavior is similar to that of the \emph{phonons} in solids, except that there are no traveling, only standing, waves; even so, these vibrations may be considered as phonons with zero mean free path because of complete disorder. Though they are preferably called \it skeletal modes\rm\, or \emph{bulk modes} in molecular parlance, both designations are used here for convenience. 

Because of the similarity, it is possible to interpret them by means of Debye's theory, which
is not, in principle, restricted to crystals, since it is based on the elastic properties of continuous media. According to Debye (see Rosenberg 1988), the upper limit of the phonon angular frequency in crystals is
\begin{equation}
\omega_{ph}=(6\pi^{2}N)^{1/3}v\,\,
\end{equation}
where $N$ is the atomic density of the solid, and $v$, the sound velocity. Typically, on Earth, $N=10^{23}-10^{24}$ at.cm$^{-3}$, and $v=2000$ m.s$^{-1}$, leading to a Debye wavelength \bf of \rm about 50 to 20$\,\mu$m, roughly defining a lower limit to the phonon spectrum. Little or no phonon continuum is therefore to be expected below this limit. For a finite sample, the higher wavelength limit is set by the size and shape of the sample and increases with its size, reaching into the millimeter range.

Moreover, Debye's elementary theory also shows that the spectral density of phonons increases as the square of their frequency, i.e. maximum density occurs at Debye's wavelength. This paradigm was found to apply \emph{approximately} to most solids. Our molecules are found to exhibit a similar behavior as soon as they become big enough. The phonon spectrum extends beyond 1000 $\mu$m in the FIR, and, for not too intricate structures, like alkane chains, clearly ends near 20 $\mu$m in the mid-IR. In general, it may overlap the redder UIBs (Unidentified Infrared Bands) and the prominent PPN bands (21 and 30-35 $\mu$m). 

In the case of the molecule of Fig.\ref{Fig:hacspec}, the frontier between phonons and mid-IR CH modes is blurred because of the structure's intricacy, but no phonons appear below 5 $\mu$m. The average mode density scales like $\lambda^{-2.6}$.

As the broadening mechanisms which determine $\Delta\nu$ are usually limited, the phonon modes  only merge into a continuum when the structure has grown to micrometric sizes. This is most likely to occur in space but can hardly be modeled with available molecular chemistry software. However, we note that, even a minor change in composition or connectivity of the structure will alter the whole spectrum, if only slightly. Therefore, \it it is possible to approach a solid-state-like continuous spectrum by considering a large number of smaller, non-identical structures.  Following this course, it is found that the multiplication of molecules of the same structural type does not alter the general trends of the spectrum.\rm

Once a reasonably large number of such molecules have been modeled, it is possible to 
assume that the spectral range up to 1000 $\mu$m, in Fig.\ref{Fig:hacspec} is indeed densely filled with representative points, and proceed to the calculation of $\alpha$. If the density of the structure is taken at 2 g.cm$^{-3}$ as above and if $N_{c}$ is the number of atoms in it, Eq. 5 then becomes

\begin{equation}
\alpha=1.75\,10^{4}\,\frac{<A>}{\Delta\nu\,N_{c},}
\end{equation}

where it is subsumed that the weight of H or other atoms is negligible. If there are enough modes within the spectral width, $\Delta\nu$, then, in Eq. 5, $A(\lambda)$ must be replaced by its sum over all modes. If $d(\lambda)$ is the spectral density of modes, then, the sum of $A/\Delta\nu$ is
 
\begin{equation}
\Sigma\frac{A}{\Delta\nu}=<A(\lambda)>\,d(\lambda)\,\lambda^{2}\,10^{-4},
\end{equation}

 remembering that $\delta\nu=-10^{4}\,\delta\lambda\,\lambda^{-2}$, where $\lambda$ is in $\mu$m \bf and $\nu$ in cm$^{-1}$.\rm In the case of Fig.\ref{Fig:hacspec}, and in first approximation, between 10 and 1000 $\mu$m, $<A(\lambda)\Longrightarrow >=29.4\,\lambda^{-1.4}$ km/mole, and  $d(\lambda)=2.84\,10^{4}\,\lambda^{-2.6}$ $\mu$m$^{-1}$.

\begin{figure}
\resizebox{\hsize}{!}{\includegraphics{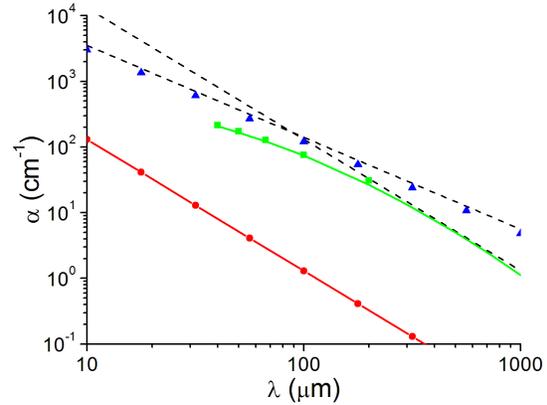}}
\caption[]{\emph{Dashed lines:} as in Fig. 5. \emph{Red dots and line}: absorption spectrum of the model a:C-H of Fig.\ref{Fig:hacspec}. \emph{Green squares and \bf line \rm}: Magnesium silicate (adapted from Day 1976) and its extrapolation to 1000 $\mu$m (shifted downward by $\times$0.103). \emph{Blue triangles}: CHONS structures (shifted downward by $\times$0.51).}
\label{Fig:alpha3}
\end{figure}

 Equation 7 finally gives $\alpha(\lambda)=1.3\,10^{4}\,\lambda^{-2}$ cm$^{-1}$ for a-C:H. Figure \ref{Fig:alpha3} shows that it is very short of the \emph{Planck} requirements; obviously, this type of structure is unsuitable for present purposes.
 
 \begin{figure}
\resizebox{\hsize}{!}{\includegraphics{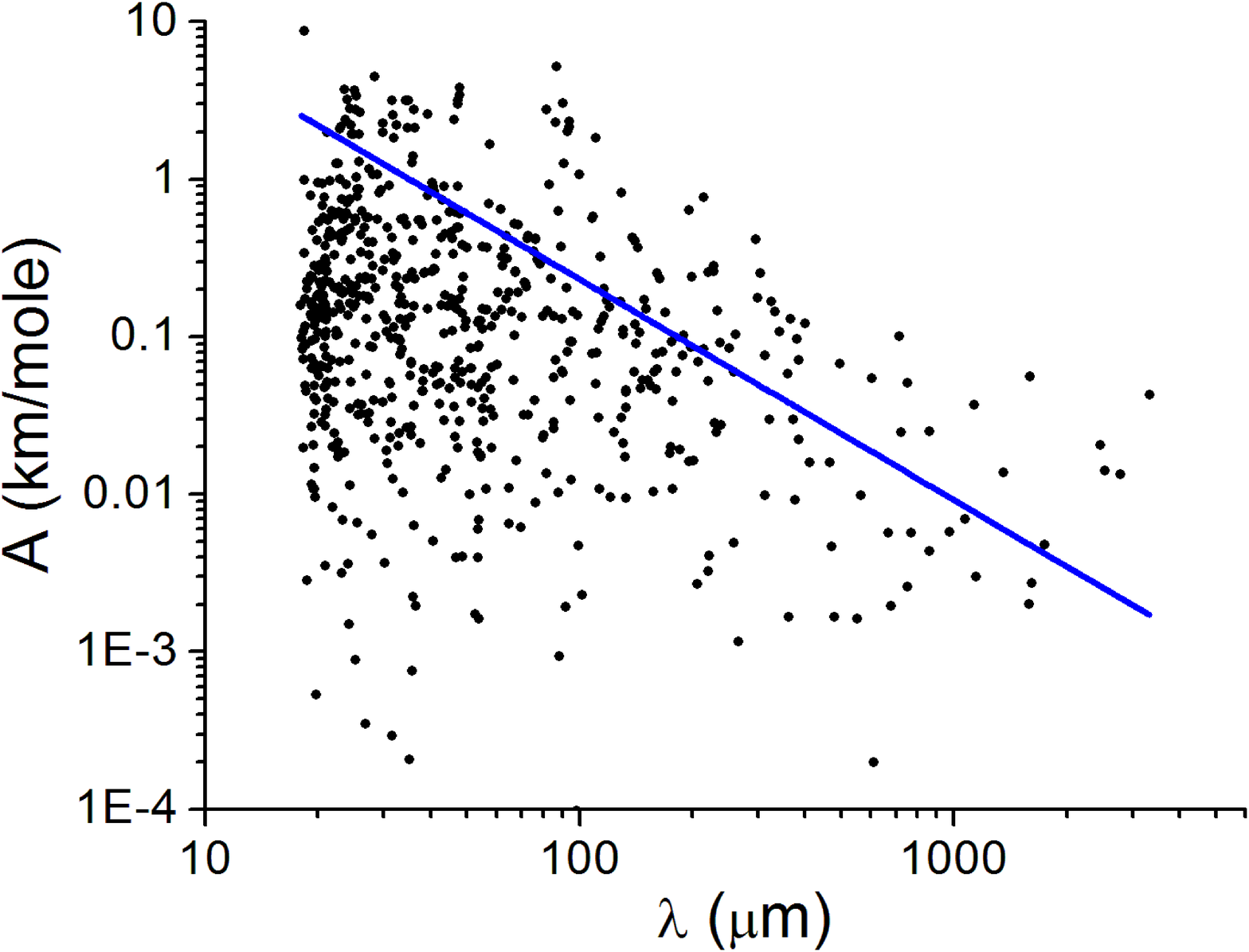}}
\caption[]{The concatenated integrated absorption phonon spectra of 21 molecules defined in Table 1. Each dot represents a vibrational mode. The line represents an average of intensities over the phonon range; its logarithmic slope is 1.4. The mode density scales like $\lambda^{-2}$.}
\label{Fig:UIBstruct}
\end{figure}

From the data base that was built as described in Papoular \cite{pap10}, it was concluded that
inclusion of aromatic rings and oxygen atoms, considerably strengthened the integrated intensities, as borne out by the measurements on laboratory HAC's (Fig.\ref{Fig:alpha2}). More complicated and richer structures than a-C:H, containing some of the most abundant elements in the ISM (H, C, O, N and S), and called CHONS for this reason, have been found in fallen meteorites and proposed as components of IS dust. On Earth, kerogens are also known to carry these elements, arranged in aromatic structures as well as pentagons, as illustrated by several proposed models reproduced in Speight \cite{spe}. These structures are found to carry the strongest integrated intensities of all the structures considered in the present study.
 
 From our data base, 21 different molecules were selected, falling in 4 categories: aromatics, aliphatic chains, oxygen-bridged chains and ``trios" (essentially a 5-membered ring squeezed between two 6-membered rings). These denominations are used here in a very restrictive sense and apply only to the present, or slightly modified, structures. Table 1 reports the compositions of the 21 molecules. Note that no N atom is included because the FIR spectra of molecules carrying nitrogen atoms do not seem to differ from those that carry silicon instead, except for the 21-$\mu$m feature observed in Planetary Nebulae, which is just outside the FIR range.

\begin{table*}[ht]
\caption[]{Atomic composition of the model CHONS dust}
\begin{flushleft}
\begin{tabular}{lllllllll}
\hline
Name & N$_{at}$ & C & H & O & S \\
\hline
Arom a & 59 & 27 & 32 & 0 & 0 \\
\hline
Arom b & 36 & 24 & 12 & 0 & 0 \\ 
\hline
Arom c & 41 & 27 & 14 & 0 & 0 \\
\hline
Arom d & 31 & 19 & 12 & 0 & 0 \\
\hline
Arom e& 44 & 22 & 22 & 0 & 0 \\ 
\hline
Aliph Ch a & 44 & 14 & 30 & 0 & 0 \\
\hline
Aliph Ch b & 53 & 17 & 36 & 0 & 0 \\
\hline
O-bridge a & 27 & 8 & 18 & 1 & 0 \\
\hline
O-bridge b & 27 & 8 & 18 & 1 & 0 \\
\hline
O-bridge c & 27 & 8 & 18 & 1  & 0\\
\hline
O-bridge d & 30 & 9 & 20 & 1 & 0 \\
\hline
O-bridge e & 33 & 10 & 22 & 1 & 0 \\
\hline
O-bridge f & 111 & 33 & 74 & 4 & 0 \\
\hline
Trio a & 57 & 30 & 26 & 1 & 0 \\
\hline
Trio b & 52 & 30 & 20 & 1 & 1 \\
\hline
Trio c & 112 & 59 & 50 & 3 & 0 \\
\hline
Trio d & 54 & 31 & 22 & 0 & 1 \\
\hline
Trio e & 52 & 30 & 20 & 1 & 1 \\
\hline
Trio f & 60 & 31 & 20 & 1 & 1 \\
\hline
Trio g & 58 & 31 & 26 & 1 & 0 \\
\hline
Trio h & 60 & 31 & 28 & 1 & 0 \\
\hline
Total & 1060 & 499 & 540 & 18 & 3 \\
\hline
\% & 1 & 47 & 51 & 1.7 & 0.3 \\
\hline
\end{tabular}
\end{flushleft}
\end{table*}

The spectrum of each molecule was computed and the power laws for density and intensity were deduced. The spectra were concatenated and added up to 3013 modes. Only the 668 modes lying beyond 18 $\mu$m are plotted in Fig. \ref{Fig:UIBstruct}. Analysis of this concatenated spectrum shows that

$<A(\lambda)\Rightarrow >=146\,\lambda^{-1.4}$  and  $d(\lambda)=1.4\,10^{4}\,\lambda^{-2}$.

As the average number of C atoms in this set of molecules is $N_{C}=24$, Eq. 7 finally gives $\alpha(\lambda)=1.49\,10^{5}\,\lambda^{-1.4}$ cm$^{-1}$, which is also plotted in  
Fig. \ref{Fig:alpha3}. As expected, the presence of aromatic rings as well as the inclusion of oxygen have raised the curve much above the a-C:H points, very near or into the error bars set by \emph{Planck}. The closeness of the CHONS absorbance to that of moderately evolved coal is not unexpected, as the component molecules of Table 1 were inspired from the numerous chemical models proposed for coals (see Speight 1994), and selected specifically for moderately aromatic coals or kerogens.

Figure \ref{Fig:alpha3} exhibits again Day's measurements on Mg$_{2}$SiO$_{4}$, extrapolated with a polynomial of order 2 into the FIR. The absorption spectral indexes of silicate and CHONS appear to make a favorable association as they bracket the $\emph{Planck}$'s preferred spectral index. We shall settle on them and seek, next, the best combination of both  material temperatures and fractional compositions to fit the observational constraints.

\section{The silicate/CHONS dust}

The function we seek to fit is 
\begin{equation}
\sigma_{0}(\frac{\lambda_{0}}{\lambda})^{\beta}BB(T,\lambda),
\end{equation}

where $T=17.9$\,K, $\beta=1.78$, $\lambda_{0}=250\,\mu$m and $\sigma_{0}=10^{-25}$ cm$^{2}$ per H atom. The fitting function is the sum of two functions of the form 

\begin{equation}
C_{m}\sigma_{m}(\lambda)BB(T_{m},\lambda),
\end{equation}

where the subscript $m$ designates the material, 1 for CHONS's and 2 for silicate, and $C_{m}$ is the corresponding fraction of the element (C or Si) that is to be present in the dust (obviously, they should be smaller than 1). Here, $\sigma_{C}=\alpha_{C}/3\,10^{26}$ and $\sigma_{Si}=\alpha_{Si}/2.9\,10^{27}$, as discussed in Sec. 2 ($\sigma$ in cm$^{2}$ per H atom, $\alpha$ in cm$^{-1}$). The standard non-linear least-squares minimization method is used to determine the 4 best-fit parameters, $C_{m}$ and $T_{m}$. They are found to be

\begin{eqnarray}
C_{1}=1.14\pm0.02, T_{1}=18.7\pm0.13 \,\mathrm{K}\\ 
C_{2}=2.45\pm0.03, T_{2}=18.56\pm0.12\,\mathrm{K}\\
\chi^{2}=8.5\,10^{-20}. 
\end{eqnarray}

\begin{figure}
\resizebox{\hsize}{!}{\includegraphics{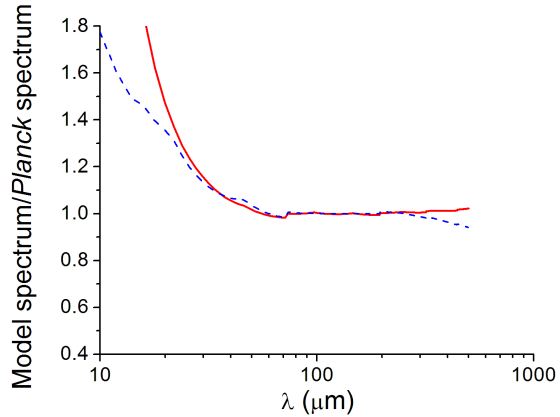}}
\caption[]{Ratio of model spectrum to \emph{Planck}'s best fit spectrum. \emph{Red line}: best fit with both free temperatures and fractions of silicate and CHONS. \emph{Blue dashes}: the same with Draine's ``astronomical" silicate and graphite. Both models require much more carbon and silicon than available in the ISM.}
\label{Fig:fit1}
\end{figure}

By analogy with Fig. 17(top) of Abergel et al. \cite{abe11}, Fig. \ref{Fig:fit1} displays
the ratio of the fitting function to the reference one (full line). The fit is seen to be much better than allowed by the observational error bars. Unfortunately, the coefficients $C$, are larger than 1; particularly so in the case of the silicate, where it is hardly possible to invoke errors in the measurement of its absorption coefficient. Figure \ref{Fig:fit1} also displays the ratio of 4-parameter best fit to reference for the case of dust composed of ``astronomical" graphite and silicate, as a dashed line (see Fig. \ref{Fig:alpha1}). In this case, the fit is slightly less good, and the parameters are again out of range:

\begin{eqnarray}
C_{1}=2.5\pm0.16, T_{1}=18.47\pm0.06\,\mathrm{K}\\
C_{2}=4.78\pm0.26, T_{2}=15.5\pm0.012\,\mathrm{K}\\
\chi^{2}=7.9\,10^{-20}. 
\end{eqnarray}

\begin{figure}
\resizebox{\hsize}{!}{\includegraphics{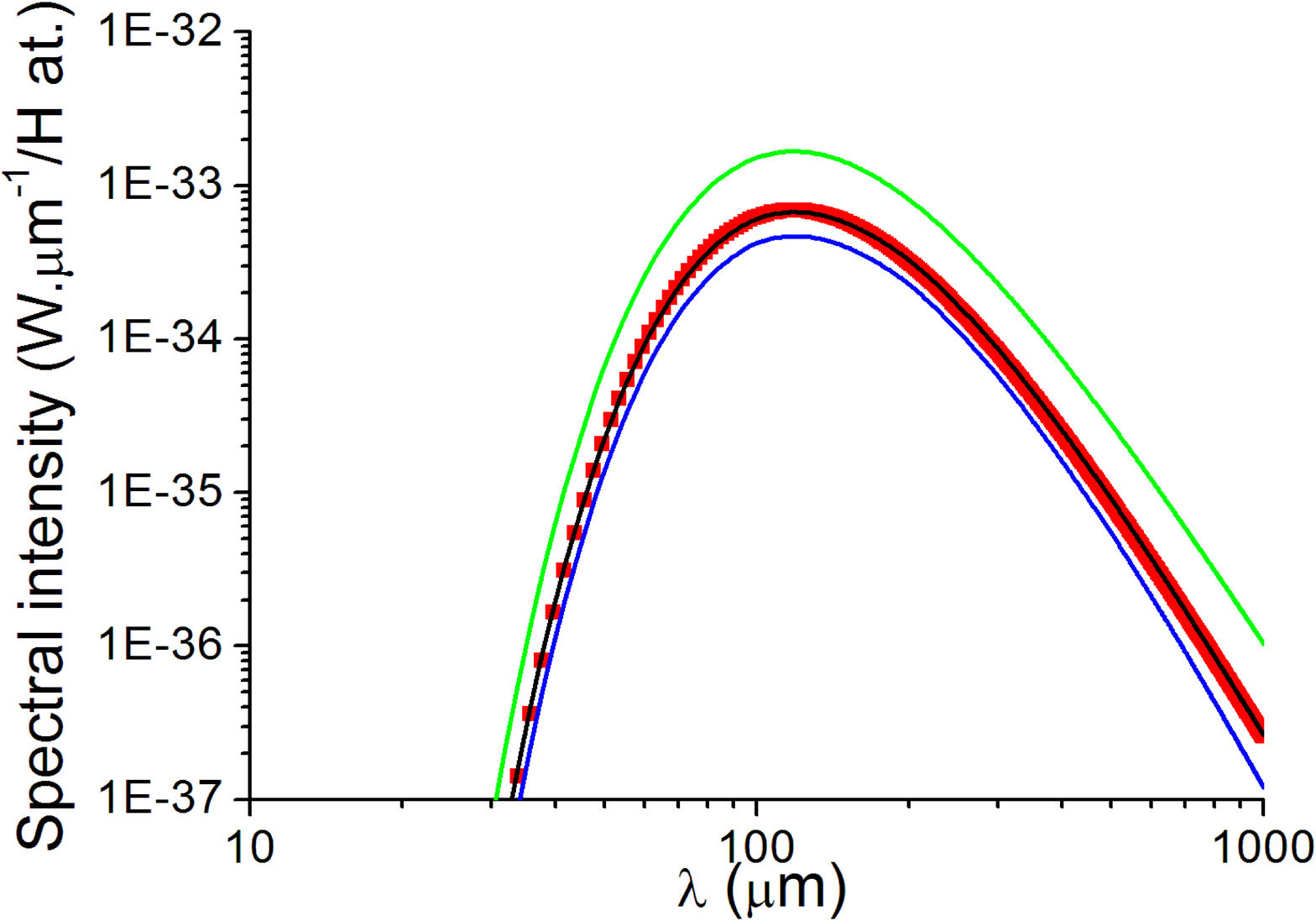}}
\caption[]{\emph{Red squares}: \emph{Planck}'s best fit spectrum. \emph{Lower blue curve}: purely silicate dust. \emph{Upper green curve}: purely CHONS dust. \emph{Intermediate black line}: present preferred model, containing 15 \% of the Galactic carbon atoms at 19 K, locked into CHONS molecules, and 90 \% of the silicate at 18.7 K; this is nearly coincident with the reference spectrum.}
\label{Fig:spectra}
\end{figure}

\begin{figure}
\resizebox{\hsize}{!}{\includegraphics{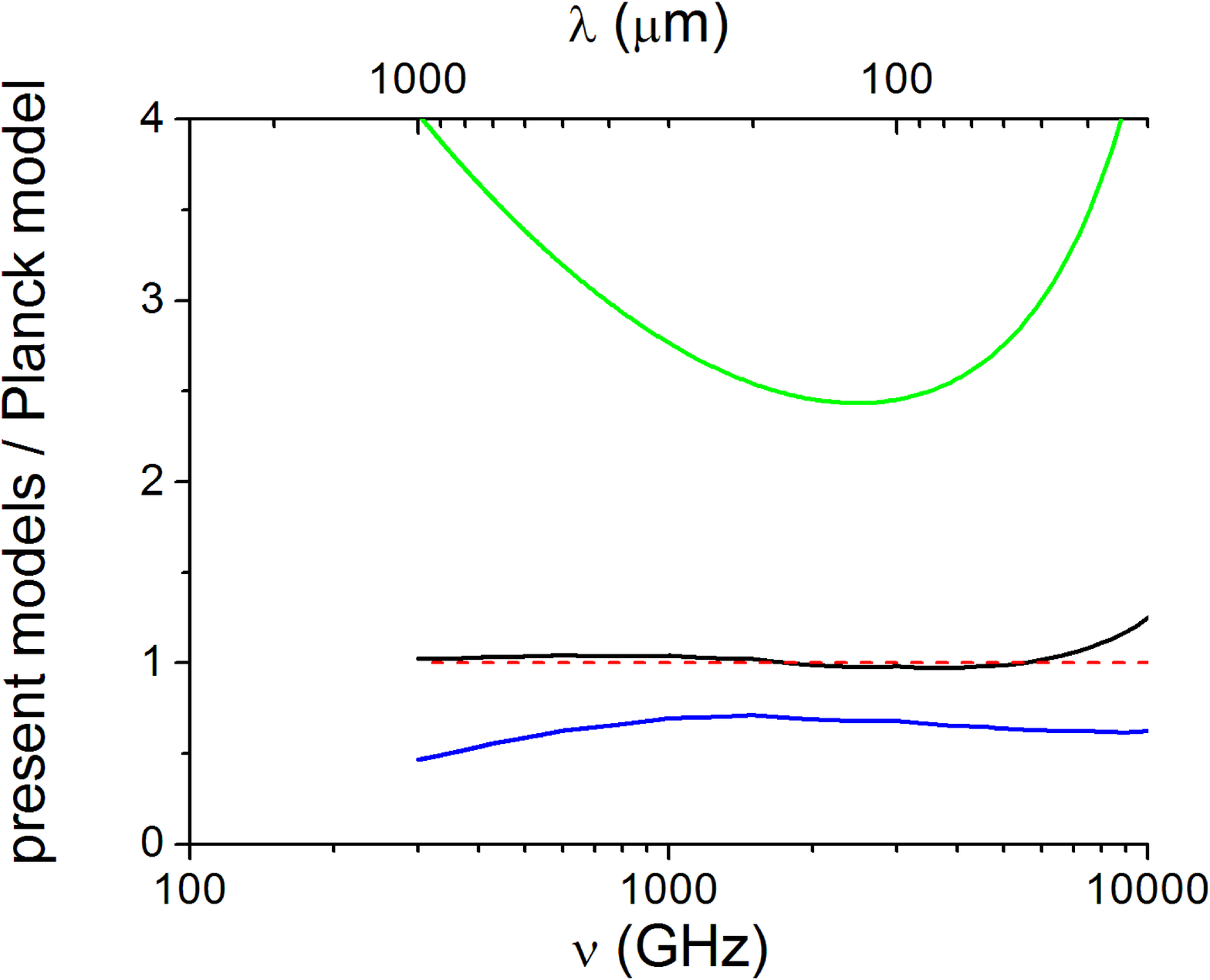}}
\caption[]{The ratio of the spectra in Fig. \ref{Fig:spectra} to \emph{Planck}'s best fit model. \emph{Lower blue curve}: purely silicate dust. \emph{Upper green curve}: purely CHONS dust. \emph{intermediate black line}: present preferred model, containing 15 \% of the Galactic carbon atoms at 19 K, locked into CHONS molecules, and 90 \% of the silicate at 18.7 K.
 The figure is to be compared with Fig. 17 (top) of Abergel et al. \cite{abe11}. Note that our preferred model falls within the error bars drawn in the latter.}
\label{Fig:modelsrat}
\end{figure}

\begin{figure}
\resizebox{\hsize}{!}{\includegraphics{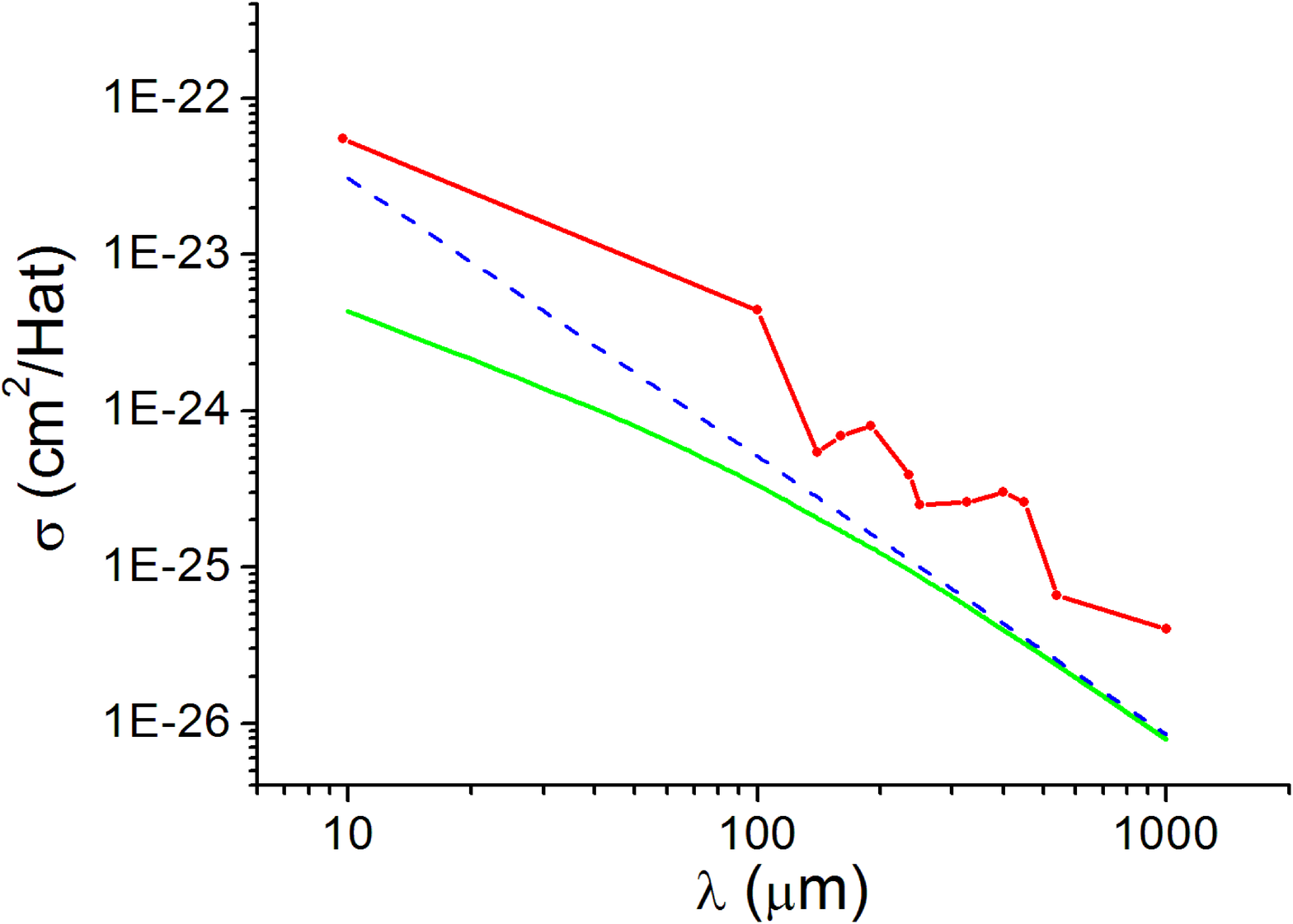}}
\caption[]{Emission cross-sections. \emph{Red dots}: Galactic measurements, adapted from Mezger et al. \cite{mez82}. \emph{Red dashes}: \emph{Planck}'s best fit model,  $\beta=1.78$ and $\sigma_{0}(250\mu$m)=$10^{-25}$ cm$^{2}$ per H atom. \emph{Blue line}: present best fit dust combination; $\sigma$(250 $\mu$m)=0.87$\,10^{-25}$ cm$^{2}$ per H atom.}
\label{Fig:sigmaGal}
\end{figure}

In view of these negative results, another fitting approach was adopted, which takes into account abundance constraints and takes fuller advantage of the complementarity of the absorption slopes of the two materials. First, the temperatures are fitted so that the peaks of the two gray spectra coincide with that of Abergel et al.'s model (Abergel et al. 2011). They are found to be 

$T(\mathrm{CHONS})=19\,\mathrm{K}; T(\mathrm{silicate})=18.7\,\mathrm{K}.$

 Note that they fall within the temperature error bars defined in Abergel et al. \cite{abe11}. The corresponding spectra are plotted in Fig. \ref{Fig:spectra}, together with $\emph{Planck}$'s reference spectrum. Then, a weighted sum of the 2 spectra is sought, which best complies with the \emph{Planck} observational error bars. Obviously, there is some leeway to such a fit. One such fit uses
 
 $C(\mathrm{CHONS})=0.15,\,\,C(\mathrm{silicate})=0.90$.

 It is the preferred model; its spectrum is also drawn, and nearly coincides with the reference spectrum. For better resolution, the ratio of the corresponding spectrum to the reference is drawn as a full line in Fig. \ref{Fig:modelsrat}; it is seen to indeed remain well within observational constraints. 
Moreover, the required fraction of Galactic C atoms, N$_{\mathrm{C}}$/N$_{\mathrm{H}}$=48 ppm, is distinctly smaller than the 60 \% which are not in the gas phase (200 ppm; Cardelli at al. 1996). For comparison, the same figure displays the curves obtained for dust containing only CHONS or  silicate, respectively. 
 
Finally, Fig. \ref{Fig:sigmaGal} displays the total absorption/emission cross-section for our preferred  model, as well as the cross-section of Abergel's model and various Galatic cross-sections previously measured and collected in Table A1 of Mezger et al. \cite{mez82}. The first two agree to a factor 2 in the observational spectral window. Both remain below the Galactic cross-sections, as they should since they only account for the FIR emission of those grains/molecules, in the high-latitude DGISM, which are large enough to reach a steady-state temperature, and then only for those Si and C atoms that are bound to the selected particular structures.

\section{Excitation and fluctuation of dust emission}

Abergel et al. \cite{abe11} plotted, for several Galactic fields, the 857 GHz (350 $\mu$m) emission as a function of HI column density (their Fig. 14). Data points``cleaned" by ``masking" cluster roughly along a straight line through the origin, which defines the average emissivity per H atom, thus confirming earlier findings. However, the contours of equal scatter point densities extend quite far away from the average line. An obvious possible cause of this dispersion is the likely patchiness of the medium density distribution, with the attendant variations in dust properties. Another may be the fluctuations of the relative abundances of carbon and silicate grains, as their spectral absorption indexes, $\beta$, are quite different: Fig. \ref{Fig:spectra} and \ref{Fig:modelsrat} suggest a possible variation of a factor 2 in peak spectral intensity. Obviously, the inception, formation and survival of these two dust components are highly dependent on ambient conditions, and could vary from site to site.

Besides, when ``masks" are removed and outlier data points are added, they are found to lie systematically, and often considerably, above the HI-correlation line (their Fig. 14; also see, Fig. 15 and 19). This FIR excess increases with $N_{\mathrm{H}}$. Several causes have been assigned to this excess. We consider here the possibility that part of it might be due to grain heating by capture of wandering H atoms. This heating agent was studied in detail in Papoular \cite{pap12}, where it was shown to be effective in the case of hydrocarbon grains with dangling carbon bonds, i.e. not capped with a hydrogen atom. This is usually the case in HI environment because of radiative or collisional dissociations, which impede full hydrogen coverage of the grain surface. In such circumstances, the capture of a H atom by a dangling bond is accompanied by an  increase of the internal energy of the grain by an amount equal to the C-H bond energy, $\sim$4.4 eV. Subject to assumptions detailed in that work, the ratio of H-capture to radiative-heating power is 

$r=0.11*n_{\mathrm{H}}/\mathrm{G},$

where $n_{\mathrm{H}}$ is the local atomic hydrogen density in cm$^{-3}$ and G, the exciting radiative flux measured in units of 10$^{9}$ 4-eV photons cm$^{-2}$s$^{-1}$. The prefactor is naturally proportional to the dangling bonds coverage of the grain surface; here this is taken as 0.5, but is likely to vary notably from site to site (thus adding another cause of dispersion). For high Galactic latitudes low velocity (local) thin clouds, G=1 and $n_{\mathrm{H}}$ may reach 100 cm$^{-3}$, giving a 10-fold advantage to H-capture excitation. At very high galactic latitudes, the effect is enhanced by the decrease of G. Papoular \cite{pap12} argued that the global Galactic FIR excess, estimated by Mezger \cite{mez78} at a factor 8, could be explained in this way. 

Note that the rate of power absorption through this process is proportional to the density of ambient H atoms and to the abundance of C atoms in the grains, which was, itself, assumed from the start to scale like that of hydrogen. This heating rate scales, therefore, as $n_{\mathrm{H}}^{2}$. From the view point of the illuminated grain, this is equivalent to an increase of G to G' with

$G'=G(1+r)$.

This effect could thus account for the increase of the residuals, $\sigma_{S}$, with $n_{\mathrm{H}}$ (Abergel et al. 2011, Fig. 15). The quadratic dependence of heating rate upon the HI density also results in the ensuing emission being sensitive to the patchiness of the medium, which may contribute to excursions from the strict linearity with HI column density, even in quiescent clouds with uniform grain optical properties.

While evolution of the grain population was overlooked above, it must be remembered that the increase of $r$ is governed by the same factor, $n_{\mathrm{H}}/\mathrm{G}$, as the fraction of molecular hydrogen in the gas, because of the chemical kinetics of dissociation/recombination (Papoular, 2012b). This suggests that dust formation may also be enhanced, which would increase  brightness. Similarly, changes in the efficiency of H-heating may change with H-coverage of carbonaceous grains, and with the velocity of ambient H atoms. Both effects could contribute to the residuals.

\section{Discussion}

The present model does not require a particular grain size distribution. However, for a dust grain to reach a steady-state temperature under excitation by the ambient radiation, it must be able to absorb at least one photon of this radiation during one radiative relaxation time. We take the average ambient photon flux to be 10$^{9}$ 5-eV-photons.cm$^{-2}$s$^{-1}$. Now, the absorption efficiency, $Q/a$, of most carbonaceous materials in the UV is of the order of 5$\,10^{5}\,$cm$^{-1}$, where $a$ is the radius of the dust grain, assumed to be spherical. The radiative relaxation time of a molecule in the FIR lies between 1 ms and 1 s, so if the grain is to absorb more than one photon before it relaxes radiatively, then $a$ should be larger than 0.1--1 $\mu$m. For an atomic density of $3\,10^{22}$ cm$^{-3}$ (that of graphite), this implies clusters of $10^{11}$ atoms, reminiscent of, for instance, the homogeneous porous aggregates and fractal clusters of Rouleau and Martin \cite{rou}. Such clusters are currently produced and experimentally studied in the laboratory. They are fluffy and sparse, and held together by van der Waals forces and hydrogen bonds. It can therefore be assumed that, in space, molecules like those studied above also aggregate upon collision over the ages to ultimately grow into clusters of the required size. Thanks to their fluffiness, their FIR optical properties should not be different than those of their mother molecules. According to Fig. \ref{Fig:alpha3}, then, the optical thickness of a 1 $\mu$m-grain at 250 $\mu$m should not much exceed $3\,10^{-3}$. This justifies \emph{a posteriori} the spectral calculations of Sec. 5.

The weak binding forces that hold the clusters together should not be able to withstand IS shocks, which may explain the decrease of the FIR luminosity observed in the higher-velocity IVC's relative to local, quiescent LVC's (Abergel et al. 2011).

On the other hand, it must be remembered, as explained in Sec. 6, that the surface of the model molecules is expected in thin space to present a considerable fraction of dangling bonds. This should favor the creation of new covalent bonds upon encounters between molecules. One must therefore consider the likelihood for the aggregated structures to be dense rather than fluffy. 
Modeling large structures of this sort far exceeds the capacity of present-day chemical software.
In that case, laboratory measurements should be undertaken instead. Macroscopic samples having the same structure as our models are readily available in the form of kerogen material. After all, weakly graphitized coal (a close relative of kerogen) was shown in Fig. \ref{Fig:alpha2} to come already quite close to observational requirements, and the trend displayed in that figure suggests that bulk kerogen, which is still less graphitized, will come still closer, as its spectral index increases slightly.

Assuming the average atomic mass of the gas is 1.13 a.m.u., and those of the CHONS and silicate clusters are respectively those of CH and Mg$_{2}$SiO$_{4}$, i.e. 13 and 140 a.m.u., then the corresponding dust-to-gas mass ratios are about 1/2063 and 1/315, for a total of 1/273.

The molecules and clusters that are too small to reach an equilibrium temperature can be shown to emit, through a different mechanism, a spectrum which differs from a gray-body spectrum (see Papoular 2012 and 2013). This spectrum extends over the phonon continuum but is weaker than the gray-body spectrum modeled above. It can therefore fill the observed emission range between the latter and the UIB spectrum i.e. about 20 to 100 $\mu$m. These particles of intermediate size may thus be considered a possible incarnation of the Very Small Grains (VSG) invoked by Desert et al. \cite{des90}, or the small amorphous carbon grains of Compiegne et al. \cite{com11}, Fig. 2. Full treatment of this subject, however, is outside the scope of the present paper.

The present study indicates that ``amorphous" carbons devoid of any aromatic component have too weak emissivity for our purposes. On the other hand, too strongly aromatic ``amorphous" carbons have too low $\beta$. Only glassy carbon, in between them, would fit. However, its awkward fibrilic structure is unlikely to form readily in space. By contrast, CHONS molecules are disordered but include a sizable fraction of aromatic rings. Their formation in space, starting from small molecules, is straightforward and they agglomerate easily into clusters. They have also been shown to carry the UIBs. Again, their $\beta$ ($\sim 1.4$) nicely matches that of silicate ($\sim 2$) to give a more fitting absorption index.

It was made clear that there is no universal value of $\beta$. The latter depends on the structure and composition of the material at hand. Here, it was found to fall mostly between 1 and 2. \emph{For a given material, it is determined by the variation with wavelength of both the vibrational mode intensities and densities}. Only as a first approximation, and in the spectral range of present interest, do these obey simple power laws in $\lambda$, with a constant spectral index. There are indications, in our data bank, of some upward bending of the average intensity beyond 1000 $\mu$m, but that is a matter for further investigation.

In conclusion, it appears that a reasonable fraction of the Galactic carbon atoms locked into CHONS clusters larger than 0.1-1 $\mu$m in radius can account for the gray-body spectrum observed by the \emph{Planck} satellite, and others before, in the FIR, if it is associated with silicate grains carrying nearly all the Galactic silicon atoms. While the ``astronomical graphite" proposed by Mathis et al. \cite{mat83} or Draine \cite{dra85} could provide equally acceptable fits, the CHONS clusters provide a more specific embodiment of the carbonaceous dust.

Note: After submitting this manuscript I noticed that the \emph{Planck} team very recently changed their preferred average parameters to $T=19.7\,$K,$\beta=1.62$ with uncertainties of order 3-6 \% (Abergel et al. 2013). This does not impact the main conclusions of the present work. 

\section{Acknowledgments}
I am grateful to the reviewer, Dr Th. Posch, for many helpful comments and suggestions.

\end{document}